\documentclass[doublecol,linenumbers]{epl2}
\usepackage{epsfig}
\usepackage{dcolumn}
\usepackage{amsmath}
\begin{document}
\title{\bf\ Theoretical model for negative giant magnetoresistance in ultra-high-mobility
2D electron systems.}

\author{J. I\~narrea}
\institute{Escuela Polit\'ecnica
Superior, Universidad Carlos III, Leganes, Madrid, 28911, Spain}
\pacs{nn.mm.xx}{First pacs description}
\pacs{nn.mm.xx}{Second pacs description}
\pacs{nn.mm.xx}{Third pacs description}

\date{\today}
\abstract{
We report on theoretical studies of the recently discovered negative giant magnetoresistance
 in ultraclean two-dimensional
electron systems at low
temperatures.
We adapt a transport model
to a ultraclean scenario  and calculate the
elastic scattering rate (electron-charged impurity) in a regime where the Landau level
width is much smaller than the cyclotron energy. We obtain that for low magnetic fields the
scattering rate and, as a consequence,  the longitudinal magnetoresistance
dramatically drop because of the small  density of states between Landau levels.
We also study the dependence of this striking effect on temperature and
an in-plane magnetic field.}

\maketitle
\section{Introduction}
Electron transport assisted by external AC or DC fields in  low-dimensional electron systems  has been always
 a central topic in basic and applied research in Condensed Matter Physics.
An important outcome  is that in the last decade the quality and hence the mobility of two-dimensional electron systems (2DES),
have been continuously increasing, exceeding routinely now the $10^{7}$ $cm^{2}/Vs$ level.
As a result of that, unexpected physical phenomena have been discovered such as,
for example, the microwave-induced mangetoresistance ($R_{xx}$) oscillations (MIRO)
and zero resistance states (ZRS). These effects
were  discovered  when a 2DES in a low and perpendicular magnetic field ($B$)
was irradiated with microwaves (MW) \cite{mani1,zudov1}.
Different
theories have been proposed to explain these  effects
\cite{ina2,ina20,girvin,dietel,lei,rivera} but the physical
origin  still remains unclear.
In the same way, a great effort has  been also made from the experimental side
\cite{mani2,mani3,willett,mani4,smet,yuan,vk,manir1,manir2}.

An interesting and challenging experimental
result, recently obtained\cite{yanhua,hatke2} and as
intriguing as ZRS,
consists in a strong resistance spike which shows up far off-resonance.
It occurs  at twice the
cyclotron frequency, $w\approx2w_{c}$\cite{yanhua,hatke2},
were $w$ is the radiation
frequency and $w_{c}$ the cyclotron frequency.
The amplitude of such a spike is
very large reaching an order of magnitude regarding MIRO.
Another remarkable result in the same experiments is
a dramatic drop in the magnetoresistance, in other words, it is obtained negative giant
magnetoresistance (NGMR) confined at low $B$, ($B\leq 0.1T$).
The appearance of this effect is concurrent with the off-resonance magnetoresistance spike and
always in very high mobility samples. In all previous experiments  about
MIRO and ZRS, using lower mobility samples, this concurrence
was never obtained.
Therefore, we must conclude that, in some way, the two physical phenomena have to be
connected or  share the same physical origin. The first experiment to obtaine NGMR, (without
irradiation) was carried out by
\revision{Paalanen et al\cite{paal} and later on by} Bockhorn et al.\cite{bock}, where they study the dependence
of NGMR on temperature and electron density.
Next,  Y. Dai et al.\cite{yanhua2}, reported on the dependence of NGMR on an in-plane $B$.
More recently Hatke et al.\cite{hatke3}, obtained experimental results on the dependence of NGMR on temperature and
a tilted $B$. Finally, the most recent experimental results on NGMR are by Mani et al., where they
report on the dependence of NGMR on the sample size\cite{mani5}.
One important outcome, common to all experiments, is that when increasing the
sample disorder, NGMR tends to progressively disappear. In this way, higher temperatures, more intense in-plane $B$ and
bigger sample size, all of them contribute to increase the disorder,  giving rise to a vanishing NGMR.
On the theoretical side, although some works have been published on the magnetoresistance spike\cite{ina3,volkov},
however no  theoretical approach
 trying  to explain  the physical origin of NGMR or its
connection with the off-resonance magnetoresistance spike, has been presented to date.

In this article, we
theoretically study and discuss the physical origin of NGMR and its dependence
on temperature and an in-plane $B$. We extend a previous transport model for 2DES based
in elastic scattering between Landau levels (LL) due to charged impurities\cite{ina2,ridley}.
This transport model was developed by the authors to deal with MIRO and ZRS\cite{ina2,ina20,ina4,ina5}.
We adapt this model to ultraclean samples,
obtaining that the scattering conditions are strongly modified. Mainly because the LL, which in
principle are broadened by scattering,  become
very narrow in this kind of samples. This implies an increasing number of states
at the center of the LL sharing a similar energy. However, in between LL, it happens the opposite,
the density of states  dramatically decreases (see Fig.1).
\begin{figure}
\centering \epsfxsize=2.8in \epsfysize=2.4in
\epsffile{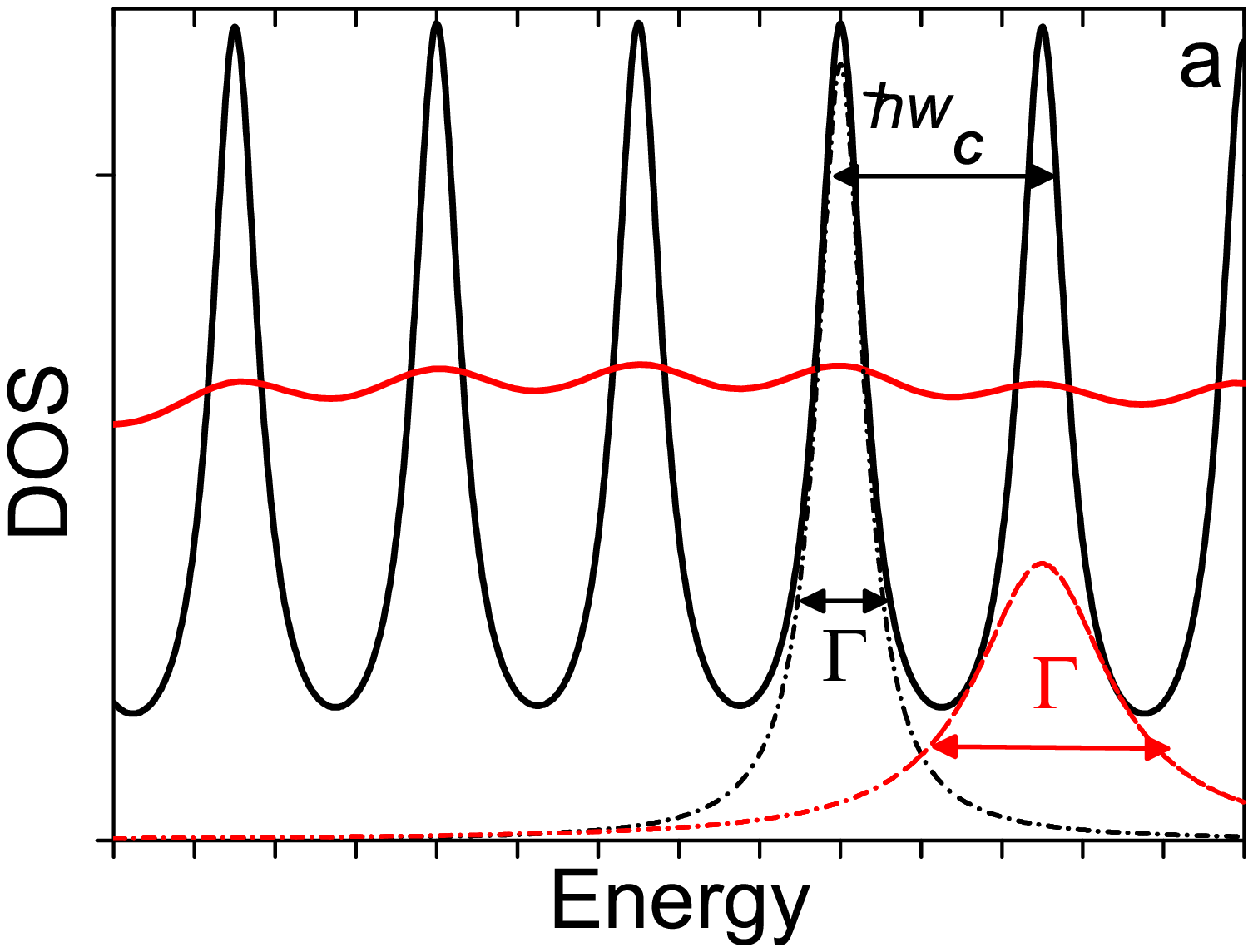}
\centering \epsfxsize=2.8in \epsfysize=2.9in
\epsffile{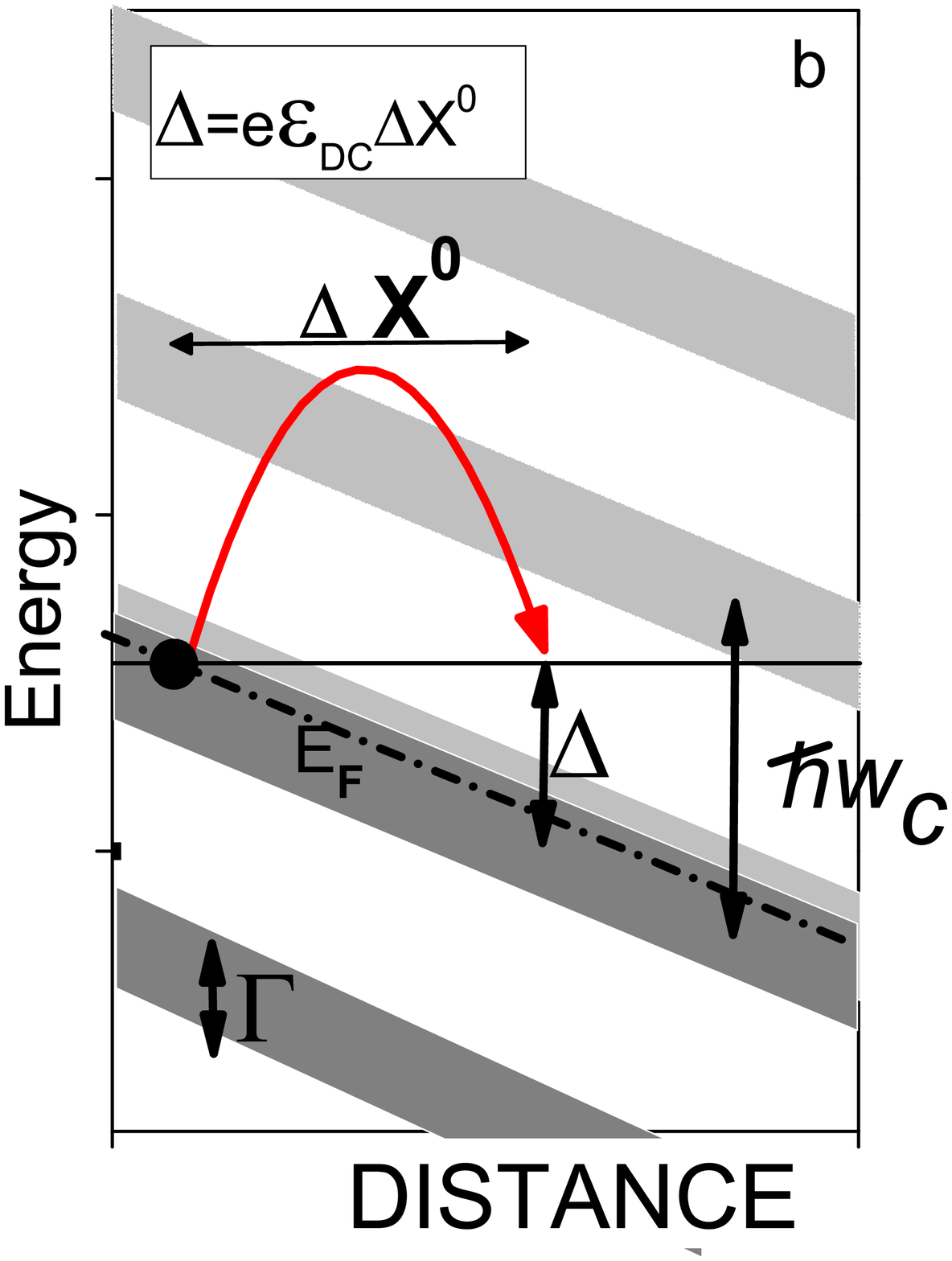}
\caption{a) Schematic diagram showing the density of Landau states simulated by Lorentzian functions
for wide and narrow Landau levels; the  width of the states is indicated by  $\Gamma$. b)
Schematic diagram presenting the elastic scattering process for ultraclean samples. The narrow Landau levels
are presented by stripes of grey color. They are tilted by the action of the static electric field
$\xi_{DC}$ in the current direction ($x$-direction). The subsequent energy drop in the scattering jump is given by $\Delta=e\xi_{DC}\Delta X^{0}$.}
\end{figure}
 We show that,  at low $B$ and for the standard DC static electric field used in these experiments, the final density of states in  the elastic scattering process,
corresponds to a region between LL, therefore with a very low density of states. This gives rise to a
small scattering rate and eventually an important drop in the measured current and $R_{xx}$.
The result of an increasing temperature or  an in-plane magnetic field is to make bigger
the disorder of the sample increasing, in turn, the LL width. As a result there will be
more available  states between LL, resulting in a stronger  scattering rate and a bigger $R_{xx}$.
The final outcome is that NGMR tends to vanish.

\section{Theoretical model}
In our model of transport we basically follow the approach by Ridley\cite{ridley} and
  calculate first the scattering suffered by the electrons due to
charged impurities (elastic) applying time dependent first order perturbation theory. Thus, we
calculate the scattering rate\cite{ina2,ina4}
between two  Landau states (LS), the initial,  $n$,  and the
final, $m$ with the Fermi's Golden Rule:
\begin{equation}
W_{I}=\frac{2\pi}{\hbar}|<\phi_{m}|V_{s}|\phi_{n}>|^{2}\delta(E_{n}-E_{m})
\end{equation}
where $\phi_{n}$ and $\phi_{m}$ are the wave functions corresponding to the initial and final LS respectively,
i.e., they represent quantum harmonic oscillators.
 $V_{s}$ is the scattering potential for charged impurities\cite{ando},
$V_{s}= \sum_{q}\frac{e^{2}}{2 S \epsilon (q+q_{s})} \cdot e^{i
\overrightarrow{q}\cdot\overrightarrow{r}}$
$S$ being the surface of the sample, $\epsilon$ the GaAs dielectric
constant, and $q_{s}$ is the Thomas-Fermi screening
constant\cite{ando,davies}.
$E_{n}=\hbar w_{c}(n+1/2)$ and $E_{m}=\hbar w_{c}(m+1/2)-\Delta$ are the corresponding LS energies for the initial
and final states respectively. $\Delta$ is the energy drop along the scattering
jump up to the final LS due to static electric field $\xi_{DC}$.
$\xi_{DC}$ is aligned with the $x$ direction and is, in turn, responsible of the current (see Fig. 1).
A  more elaborated expression for the
scattering rate can be obtained\cite{ina2,ridley} being given by:
\begin{eqnarray}
W_{I}&=& \frac{e^{4}n_{i}}
{8\pi^{2}\hbar \epsilon^{2}}\sum_{m}\left[\frac{\Gamma}{[E_{n}-E_{m}]^{2}+\Gamma^{2}}\right] \nonumber\\
&&\times\int d\theta \int dq
\frac{q}{(q+q_{s})^{2}} e^{-\frac{1}{2}q^{2}R^{2}}
\left[L_{n}\left(\frac{1}{2}q^{2}R^{2}\right)\right]^{2}
\end{eqnarray}
where $n_{i}$ is the impurity density, $R$ is the magnetic
characteristic length, $R^{2}=\frac{\hbar}{eB}$.
$L_{n}$ are the associated Laguerre polynomials.
In the obtained  expression for the impurity scattering rate, the delta function, $\delta(E_{n}-E_{m})$, has
been approached  by a Lorentzian,
considering that the LL are
broadened by disorder,
\begin{equation}
\delta(E_{n}-E_{m})\simeq \frac{1}{\pi}\frac{\Gamma}{(E_{n}-E_{m})^{2}+\Gamma^{2}}
\end{equation}
where $\Gamma$ is the LL width.
On the other hand, the sum is carried out up to all final LS, ($\sum_{m=0}^{\infty}$).

When it comes to extending the transport model to a ultraclean scenario it is
essential to consider that now the LL width is much smaller than $\hbar w_{c}$.
Accordingly, we first apply the Poisson sum rules to perform the infinite sum of LL in eq. (2) and obtain:
\begin{eqnarray}
&&\sum_{m}\left[\frac{\Gamma}{[E_{n}-E_{m}]^{2}+\Gamma^{2}}\right]=\nonumber\\
&&=\frac{1}{ \hbar w_{c}}\Bigg\{1+2\sum_{s=1}^{\infty} \cos\left[ \frac{2\pi s \Delta}{\hbar w_{c}} \right]
e^{\left[-\frac{\pi\Gamma s}{\hbar w_{c}}\right]} \Bigg\}
\end{eqnarray}
 When $\Gamma\ll \hbar w_{c}$, (ultra-clean scenario), is highly recommended, if possible, to carry out  the total sum over $s$  inside
 the curly brackets\cite{grads}:
\begin{eqnarray}
&&\sum_{s=1}^{\infty}\left[\cos\left[\frac{2\pi s \Delta}{\hbar w_{c}} \right]exp\left[-\frac{\pi\Gamma s}{\hbar w_{c}}\right]\right]=\nonumber\\
&&\nonumber\\
&&=\frac{e^{\left[-\frac{\pi\Gamma }{\hbar w_{c}}\right]}  \big\{  \cos \left[\frac{2\pi  \Delta}{\hbar w_{c}}\right]- e^{\left[-\frac{\pi\Gamma }{\hbar w_{c}}\right] }\big\}}
{1-2e^{\left[-\frac{\pi\Gamma }{\hbar w_{c}}\right]}   \cos \left[\frac{2\pi  \Delta}{\hbar w_{c}}\right]+e^{\left[-\frac{2\pi\Gamma }{\hbar w_{c}}\right]} }
\end{eqnarray}
to  finally obtain,
\begin{eqnarray}
&&\sum_{m}\left[\frac{\Gamma}{[E_{n}-E_{m}]^{2}+\Gamma^{2}}\right]=\nonumber\\
&&\frac{1}{ \hbar w_{c}}\Bigg\{ \frac{1-e^{\left[-\frac{2\pi\Gamma }{\hbar w_{c}}\right]} } {1-2e^{\left[-\frac{\pi\Gamma }{\hbar w_{c}}\right]}
\cos \left[\frac{2\pi  \Delta}{\hbar w_{c}}\right]+e^{\left[-\frac{2\pi\Gamma }{\hbar w_{c}} \right]} }  \Bigg\}
\end{eqnarray}
\\

 Then, substituting this result
 into $W_{I}$ we get to:
\begin{equation}
W_{I}\propto\Bigg\{ \frac{1-e^{\left[-\frac{2\pi\Gamma }{\hbar w_{c}}\right]} } {1-2e^{\left[-\frac{\pi\Gamma }{\hbar w_{c}}\right]}   \cos \left[\frac{2\pi  \Delta}{\hbar w_{c}}\right]+e^{\left[-\frac{2\pi\Gamma }{\hbar w_{c}} \right]} }  \Bigg\}
\end{equation}
Once we know the  scattering rate, we consider that  when
an electron undergoes a scattering process, due to charged impurities,  its average orbit center position
changes in the static electric field direction, ($\xi_{DC}$ or $x$ direction), from $X_{0}$ to $X_{0}^{'}$.
Accordingly, it advances an average effective distance given by\cite{ina2,ina3}:
$ \Delta X^{0}=X_{0}^{'}-X_{0}\simeq 2R_{c}$\cite{du0},
$R_{c}$ being  the orbit radius $R_{c}=\sqrt{2m^{*}E_{F}}/eB$.
Therefore, $\Delta$ and $\xi_{DC}$ are related by $\Delta=e\xi_{DC}\Delta X^{0}$.

Now, we can obtain the expression
for the longitudinal conductivity $\sigma_{xx}$ according
to\cite{ina2,ina20}:
$\sigma_{xx}\propto \int dE \frac{\Delta X^{0}}{\tau}$
being $E$
the energy and $\tau=\frac{1}{W_{I}}$ the scattering time.
To obtain $R_{xx}$ we use
the relation
$R_{xx}=\frac{\sigma_{xx}}{\sigma_{xx}^{2}+\sigma_{xy}^{2}}
\simeq\frac{\sigma_{xx}}{\sigma_{xy}^{2}}$, where
$\sigma_{xy}\simeq\frac{n_{e}e}{B}$, being $n_{e}$ the electron density,  and $\sigma_{xx}\ll\sigma_{xy}$.
Finally, the expression of $R_{xx}$ reads:
\begin{equation}
R_{xx}\propto\Bigg\{ \frac{1-e^{\left[-\frac{2\pi\Gamma }{\hbar w_{c}}\right]} } {1-2e^{\left[-\frac{\pi\Gamma }{\hbar w_{c}}\right]}   \cos \left[\frac{2\pi  \Delta}{\hbar w_{c}}\right]+e^{\left[-\frac{2\pi\Gamma }{\hbar w_{c}} \right]} } \Bigg\}
\end{equation}

Thus, $R_{xx}$ directly  depends 
 on $\Gamma$ and $\xi_{DC}$. For the experimental parameters that we
are dealing with\cite{bock,yanhua2,hatke3}, we have estimated that $\xi_{DC}\sim 1-2$ V/m. On the other hand for low $B\sim0.04-0.05$ T,
 we have obtained that, in average, in
the advanced distance corresponding to a scattering jump, the LL are tilted  an energy $\Delta\simeq 3.10^{-5}$ eV. For these small $B$ the
cyclotron energy, $\hbar w_{c}\sim 7-8 \times 10^{-5}$ eV. Then, comparing both numerical values ($\frac{\Delta}{ \hbar w_{c}}\sim \frac{1}{2}$),
 we can conclude that
in ultraclean samples and low $B$, the average scenario is the corresponding
to an electron "landing", after a scattering event, between LL where there is a low density of states (see Fig. 1.b). The result is a
dramatic drop  at low $B$ in $R_{xx}$ as obtained in the experiments. Thus, when it
is fulfilled that, $\frac{\Delta}{ \hbar w_{c}}\sim \frac{1}{2}$, the first term between brackets tends to:
\begin{equation}
\Bigg\{ \frac{1-e^{\left[-\frac{2\pi\Gamma }{\hbar w_{c}}\right]} } {1-2e^{\left[-\frac{\pi\Gamma }{\hbar w_{c}}\right]}   \cos \left[\frac{2\pi  \Delta}{\hbar w_{c}}\right]+e^{\left[-\frac{2\pi\Gamma }{\hbar w_{c}} \right]} }  \Bigg\}  \rightarrow \Bigg\{ \frac{1-e^{\left[-\frac{\pi\Gamma }{\hbar w_{c}}\right]} } {1  +e^{\left[-\frac{\pi\Gamma }{\hbar w_{c}} \right]} }  \Bigg\}
\end{equation}
and if, in addition to that, $\Gamma \ll \hbar w_{c}$, for low $B$ the resulting term
decreases very much affecting $R_{xx}$ which becomes also very small, producing the effect of NGMR.
However, when increasing further $B$, it turns out that, $\frac{\Delta}{ \hbar w_{c}}\rightarrow 0$, and then:
\begin{equation}
\Bigg\{ \frac{1-e^{\left[-\frac{2\pi\Gamma }{\hbar w_{c}}\right]} } {1-2e^{\left[-\frac{\pi\Gamma }{\hbar w_{c}}\right]}   \cos \left[\frac{2\pi  \Delta}{\hbar w_{c}}\right]+e^{\left[-\frac{2\pi\Gamma }{\hbar w_{c}} \right]} }  \Bigg\}\rightarrow \Bigg\{ \frac{1+e^{\left[-\frac{\pi\Gamma }{\hbar w_{c}}\right]} } {1-e^{\left[-\frac{\pi\Gamma }{\hbar w_{c}} \right]} }  \Bigg\}
\end{equation}
\\
and  $R_{xx}$ tends to increase with increasing $B$.

In a non-ultraclean sample where $\Gamma\geq \hbar w_{c}$, the sum over
final LS can be written as:
\begin{equation}
 \sum_{m}\left[\frac{\Gamma}{[E_{n}-E_{m}]^{2}+\Gamma^{2}}\right]\simeq \frac{1}{ \hbar w_{c}}\Bigg\{1+2 \cos\left[ \frac{2\pi \Delta}{\hbar w_{c}} \right]
e^{\left[-\frac{\pi\Gamma }{\hbar w_{c}}\right]} \Bigg\}
\end{equation}
where the cosine term is strongly damped by the exponential,
preventing the appearance  of NGMR.

\section{Results}
\begin{figure}
\centering\epsfxsize=3.4in \epsfysize=3.2in
\epsffile{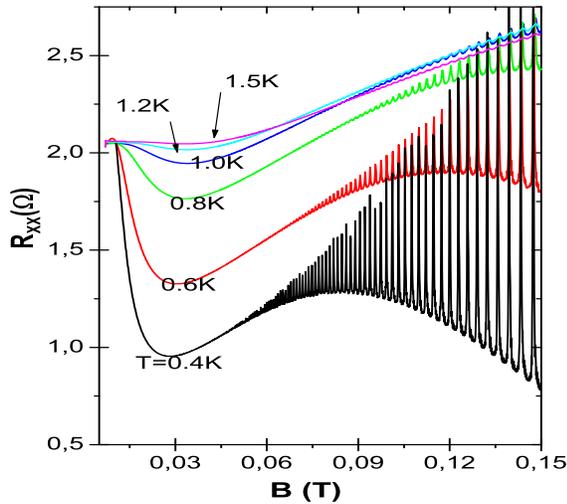}
\caption{Calculated longitudinal magnetoresistance, $R_{xx}$, vs magnetic field for several
temperatures ranging from $0.4$ K to $1.5$ K.
It is clearly observed the vanishing effect of the increasing temperature on the NGMR.}
\end{figure}
\begin{figure}
\centering \epsfxsize=3.in \epsfysize=2.0in
\epsffile{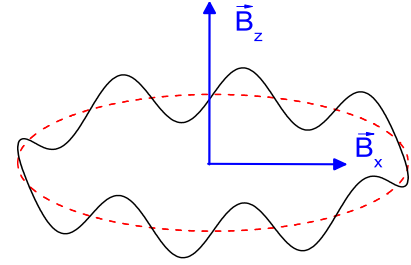}
\caption{Schematic diagram showing the semiclassical
description of electron trajectories in 2D systems in the
presence of a perpendicular $B$,($B_{z}$), and a parallel $B$,($B_{x}$).}
\end{figure}
\begin{figure}
\centering \epsfxsize=3.4in \epsfysize=3.2in
\epsffile{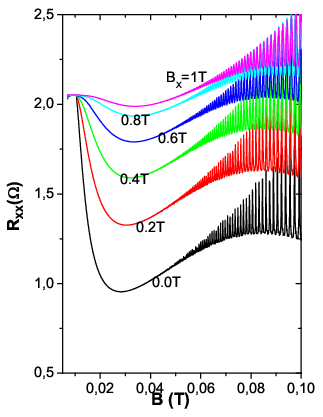}
\caption{Calculated longitudinal magnetoresistance, $R_{xx}$, vs magnetic field for different values
of the in-plane magnetic field,($B_{x}$). $B_{x}$ is ranging from
0 T to 1 T.}
\end{figure}
As we said above, the effect of both, temperature and an in-plane $B$ is to increase the disorder of the sample
with the subsequent increase of  $\Gamma$. The bigger the disorder, the wider $\Gamma$, increasing the density of states between
LL. The outcome is a stronger elastic scattering rate and eventually an increasing $R_{xx}$ and vanishing NGMR.
In the case of an increasing temperature, electrons are able to interact more strongly with the ions in the lattice,
 giving rise to a stronger  emission of acoustic phonons and  scattering rate.
This has to be
reflected in the total quantum scattering rate $1/\tau_{0}$, that encompasses all scattering sources.
According to the Matthiessen rule the total scattering rate can be
expressed as the sum of the different individual scattering sources,
$\frac{1}{\tau_{0}}=\sum_{i}\frac{1}{\tau_{i}}$
and obviously one of them is the acoustic phonon scattering rate.
Then, an increase in the phonon scattering
rate ($\gamma_{ac}=1/\tau_{ac}$) due to temperature will eventually make the total
scattering rate to increase too. According to Ando et al\cite{ando} $\gamma_{ac}$ depends
linearly on $T$:  $\gamma_{ac} \propto T$ and
then $\Delta \gamma_{ac}\propto \Delta T$. This will be reflected in
the final LL width, $\Gamma_{f}$, that can be expressed as: $\Gamma_{f}=\Gamma_{i}+\hbar \Delta \gamma_{ac}$, where
$\Gamma_{i}$ is the initial LL width corresponding to the initial temperature.
The effect of temperature is presented in Fig. 2, which exhibits $R_{xx}$ as a function of
$B$ for several temperatures ranging from 0.4 K to 1.5 K. It is clearly
observed that an increasing temperature
makes NGMR to progressively disappear. When reaching $T>1$ K, NGMR is
totally wiped out.

The effect of an in-plane magnetic field, ($B_{x}$), on the
transport in a 2DES was studied  in experiments on   radiation-induced
resistance oscillations\cite{du,mani6}.
Experimental results showed that the main effect was a progressive damping of
the whole resistance response as $B_{x}$ increased.
Subsequent theoretical results\cite{inainplane}, confirmed and
explained the surprising damping in the framework of the
radiation-driven electrons orbits model: the presence of $B_{x}$
imposes an extra harmonically oscillating motion in the
$z$-direction enlarging the electrons trajectory in their cyclotron orbits (see Fig. 3).
This would increase the interactions of electrons with the lattice and with
the walls of the quantum well giving rise of a stronger
emission of acoustic phonons. Therefore, the effect of the presence of  $B_{x}$
 is to increase the disorder in the sample and the width of the LL.
The relation between $\gamma_{ac}$ and $B_{x}$ is given by\cite{inainplane}:
\begin{equation}
\gamma_{ac}= \gamma_{ac}(B_{x}=0)  \times
\sqrt{1+\left(\frac{eB_{x}z_{0}^{2}}{\hbar}\right)^{2}}
\end{equation}
where $z_{0}$ is the effective length of the electron
wave function when we consider a parabolic potential for the $z$-confinement\cite{davies,ando}.
Now, proceeding similarly as before with temperature, we can express the
final width of LL as, $\Gamma_{f}=\Gamma_{i}(B_{x}=0)+\hbar \Delta \gamma_{ac}$,
where in this case $ \Delta \gamma_{ac}= \gamma_{ac}(B_{x}\neq0)-\gamma_{ac}(B_{x}=0)$.
The effect of  $B_{x}$ is presented in Fig. 4, where we exhibit  $R_{xx}$ as a function of
$B$ for several $B_{x}$ ranging from 0. T to 1.0 T. We observe that for $B_{x}\simeq 1 T$,
 NGMR totally disappears.
 It has been also recently reported by Mani et al.\cite{mani5} on the effect of sample size on NGMR . They
 report that NGMR is more pronounced in smaller samples and that the effect progressively disappears as
the sample size increases. The explanation can be readily obtained, at least qualitatively,  with similar terms as temperature and
$B_{x}$. In this case smaller samples present in average a weaker scattering and the
electron transport  gets closer to quasi-ballistic. Therefore, in this kind of samples $\Gamma$ will be much smaller
presenting a clear NGMR. Increasing the sample size it is expected that $R_{xx}$ will increase and NGMR will disappear,
as experiments report.

\section{Conclusions}
In summary, we have reported, from a theoretical  approach, on the recently discovered NGMR
 in ultraclean 2DES.
We adapt a transport model
to high mobility samples  and calculate the
elastic scattering rate  in a regime where the Landau level
width is much smaller than the cyclotron energy. We obtain that for low $B$ the
scattering rate and $R_{xx}$
dramatically drop because  there are very few available states where to get to.
We also study the dependence of this striking effect on temperature and
an in-plane magnetic field, and conclude that both of them  increase the disorder
of the sample giving rise to a bigger $\Gamma$ and stronger scattering rate.
The subsequent results is a greater  $R_{xx}$ and a vanishing NGMR.




\acknowledgments
This work is supported by the MCYT (Spain) under grant
MAT2011-24331 and ITN Grant 234970 (EU).

\end{document}